\documentclass[twocolumn,showpacs,preprintnumbers,amsmath,amssymb]{revtex4-1}
\usepackage{graphicx}
\usepackage{textcomp}

\begin{document}
\title{Transparency in nonlinear frequency conversion}
  \normalsize
\author{Stefano Longhi}
\address{Dipartimento di Fisica, Politecnico di Milano and Istituto di Fotonica e Nanotecnologie del Consiglio Nazionale delle Ricerche, Piazza L. da Vinci
32, I-20133 Milano, Italy}

%
\bigskip
\begin{abstract}

Suppression of wave scattering and the realization of transparency effects in engineered optical media and surfaces have attracted great attention in the past recent years. In this work the problem of transparency is considered for optical wave propagation in a nonlinear dielectric medium with second-order $\chi^{(2)}$ susceptibility. Because of nonlinear interaction, a reference signal wave at carrier frequency $\omega_1$  can exchange power, thus being amplified or attenuated,  
when phase matching conditions are satisfied and frequency conversion takes place. Therefore, rather generally  the medium is not transparent to the signal wave because of 'scattering' in the frequency domain. Here we show that broadband transparency, corresponding to the full absence of frequency conversion in spite of phase matching, can be observed for the signal wave in the process of sum frequency generation whenever the effective susceptibility $\chi^{(2)}$ along the nonlinear medium is tailored following a suitable spatial apodization profile and the power level of the pump wave is properly tuned. While broadband transparency is observed under such conditions, the nonlinear medium is not invisible owing to an additional effective dispersion for the signal wave introduced by the nonlinear interaction.

  \noindent
\end{abstract}


\maketitle

\section{Introduction}
In the past decade considerable research efforts have
been devoted in developing synthetic materials appropriately
engineered to mold the flow of light in unprecedented ways, opening the way to 
several important applications. A noteworthy example is provided by the possibility to suppress 
wave reflection and scattering from inhomogeneities or surfaces in engineered optical media (see, for instance, \cite{r0,r1,r1bis,r2,r3,r4,r5} and reference therein). 
Optical waves propagating in linear but inhomogeneous media generally experience reflection and scattering when the material properties rapidly change over a distance of the order of the optical wavelength \cite{r7bis}. In
one-dimensional purely dielectric systems, wave scattering suppression can be achieved by 
tailoring the optical refractive index to  realize reflectionless potentials. For
continuous media, the synthesis of reflectionless potentials was investigated in a pioneering work
by Kay and Moses in 1956 \cite{Kay}, and then studied in great
detail in the context of the inverse scattering theory
\cite{inv1,inv2} and supersymmetric quantum mechanics \cite{susy1}, with applications to e.g. broadband omnidirectional antireflection coatings \cite{coat} and transparent optical intersections \cite{inters}.  Exploiting the imaginary part of the dielectric permittivity $\epsilon$ (i.e. absorption) in addition to its real part, unidirectional antireflection can be also realized \cite{r5}.
In the electromagnetic domain, the full access to four quadrants of the
real $\epsilon-\mu$ plane by means of sub-wavelength structured metamaterials \cite{meta1,meta2}, in connection with methods inspired by transformation optics \cite{tra}, has widely
 extended the possibilities of controlling and suppressing wave scatting, with the demonstration of amazing phenomena like metamaterial cloaking and invisibility (for recent reviews in this broad research field see, for instance, \cite{r3,meta2,reviewmeta}).\par
In this work we consider the problem of transparency of optical waves that propagate in a nonlinear dielectric medium with second-order $\chi^{(2)}$ susceptibility. Because of nonlinear interaction, waves at different carrier frequencies  can exchange power and, when phase matching conditions are satisfied, frequency conversion generally occurs \cite{Boyd}. In such a medium 'scattering' can be viewed in 'frequency' domain rather than in the spatial one. It is well-known that nonlinear interaction of light waves in a quadratic nonlinear crystal can be exploited to properly control the spectral transmission (both in amplitude and phase) of a given reference wave at carrier frequency $\omega_1$  (signal wave). For example, in a suitably-designed optical parametric amplifier it was shown \cite{referee1} that a narrow transparency window for the signal wave can be opened, leading to superluminal group velocities. Such a narrow transparency effect, associated to superluminal propagation, basically reproduced the gain-assisted transparent pulse propagation experiment by Wang {\it et al.} \cite{referee2} in atomic vapours and shares certain similarities with electromagnetically-induced transparency (EIT). The transparency windows that can be opened in a parametric down-conversion process as well as in EIT media, however, is rather narrow. An open question is whether broadband transparency can be realized in a nonlinear optical interaction process. Here we show that, while broadband transparency can not be observed in parametric amplification, it can arise (theoretically with an infinite bandwidth) in an {\it up-conversion} process, namely in sum frequency generation (SFG) \cite{Boyd} [Fig.1(a)]. To observe broadband transparency in SFG,  the effective susceptibility $\chi^{(2)}$ along the nonlinear crystal has to be suitable apodized, which can be realized using well-established quasi-phase-matching (QPM) methods \cite{QPM,QPM2}. While broadband transparency is observed under such conditions, the nonlinear medium is not invisible, since the nonlinear interaction introduces an effective additional dispersion (phase delay) for the signal wave that can be detected by nonlinear-induced pulse distortion in a transmission experiment. 

\begin{figure}
\includegraphics[scale=0.4]{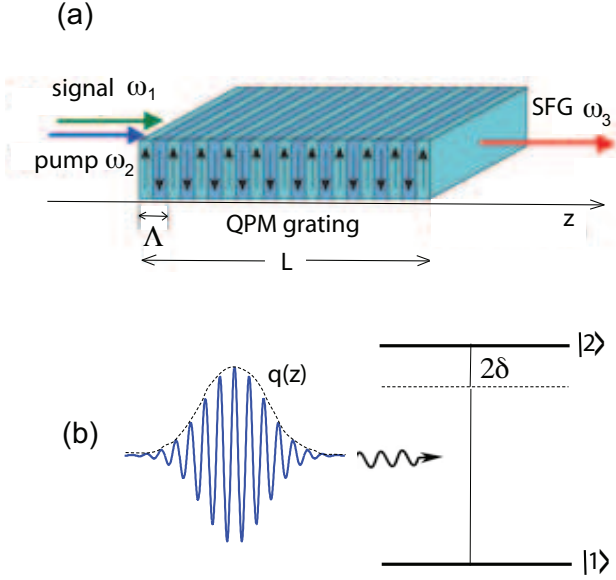}
\caption{(Color online) (a) Schematic of sum frequency generation (SFG) in a nonlinear $\chi^{(2)}$ crystal with a periodic QPM grating. A weak signal field at frequency $\omega_1$ interacts with a strong pump field at frequency $\omega_2$ to generate a SFG wave at frequency $\omega_3=\omega_1+\omega_2$. $L$ is the crystal length, $\Lambda$ is the QPM grating period. For first-order QPM grating $\Lambda= 2 \pi / \Delta k$, where $\Delta k =| k_3-k_2-k_1|$ is the wave vector mismatch of the three interacting waves. (b) Coherently-driven two-level atom analogue of the SFG process. The Rabi frequency $q(z)$ of the exciting optical pulse, with carrier frequency detuned by $2 \delta$ from the atomic transition resonance, corresponds to the apodization profile of the QPM grating.}
\end{figure}
\section{Sum frequency generation: basic equations and the driven two-level atom analogue}
\subsection{The model}
We consider parametric interaction of three co-propagating waves with carrier frequencies $\omega_1$ (signal wave), $\omega_2$ (pump wave) and $\omega_3=\omega_1+\omega_2$ (SFG wave) in a nonlinear medium of length $L$ with a second-order $\chi^{(2)}$ nonlinearity, which are phase-matched via a QPM grating [Fig.1(a)]. The  pump field at frequency $\omega_2$ is assumed to be a strong and continuous-wave field, whereas the signal at carrier frequency $\omega_1$ injected into the medium as well as the SFG wave are assumed weak but arbitrarily broadband. In the effective plane-wave
approximation and taking into account material dispersion, from Maxwell's equations the electric field $\mathcal{E}(z,t)$ is the medium is found to satisfy the nonlinear and dispersive wave equation (see, for instance, \cite{Newell92})
\begin{equation}
\frac{\partial^2 {\cal E}}{\partial z^2}+ \int_{-\infty}^{\infty} d \omega k^2(\omega)
\tilde {\cal E}(z,\omega) \exp(-i \omega t) =\mu_0 \frac{\partial^2 {\cal
P}^{NL}}{\partial t^2} \; ,
\end{equation}
where $\tilde {\cal E}(z,\omega)=(2 \pi)^{-1} \int_{-\infty}^{\infty} d\omega {\cal
E}(z,t) \exp(i \omega t)$ is the Fourier transform of ${\cal E}(z,t)$,
$k(\omega)=(\omega / c_0) \sqrt {1 + \tilde \chi(\omega)}=(\omega/c_0)n(\omega)$ is the
dispersion relation defined by the complex linear susceptibility $\tilde {\chi}
(\omega)$ [or by the complex refractive index $n(\omega)= \sqrt{1+\tilde
{\chi}(\omega)}$], $c_0$ is the speed of light in vacuum, $\mu_0$ is the vacuum
magnetic permeability, and ${\cal P}^{NL}$ is the nonlinear driving polarization term.
For a quadratic medium and neglecting dispersion and absorptive effects of second-order polarization,
one can take ${\cal P}^{NL}(z,t)=\epsilon_0 \chi^{(2)}(z) {\cal E}^2(z,t)$, where
$\chi^{(2)}$ is the spatially-modulated nonlinear susceptibility that accounts for the
QPM grating. To study the process of SFG, the electric field ${\cal E}(z,t)$
is assumed to be given by the superposition of three wave trains with carrier frequencies
$\omega_1$ (signal field), $\omega_2$ (pump field) and $\omega_3=\omega_1+\omega_2$ (SFG field), co-propagating along the longitudinal $z$ direction.  
Phase matching is accomplished by a QPM grating, i.e. the susceptibility $\chi^{(2)}$ is a quasi-periodic
function of $z$ with period $\Lambda$
\begin{equation}
\chi^{(2)}(z)= \sum_{n=-\infty}^{\infty} \chi^{(2)}_{n}(z) \exp(-2 i n \pi z / \Lambda)
\; ,
\end{equation}
where the Fourier coefficients ${\chi^{(2)}_{n}(z)}$ are slowly varying functions of
$z$ over one period $\Lambda$. In practice, the slow dependence of coefficients on $z$
can be achieved by a +/- reversal of domains in the ferroelectric crystal with a local
period and local duty cycle that are slowly varying along the $z$ axis; methods to apodize the QPM grating are described and demonstrated, for
instance, in \cite{QPM}.  For first-order QPM, the grating period $\Lambda$ satisfies the condition 
$\Lambda=2 \pi / \Delta k$, where $\Delta k \equiv |k_3-k_2-k_1|$ is the wave vector mismatch of interacting waves and $k_l=k(\omega_l)$ ($l=1,2,3$). 
After setting
\begin{eqnarray}
{\cal E}(z,t) & = &  \frac{1}{2}\left\{ A_1(z,t) \exp(-i \omega_1 t+i k_1z) \nonumber \right. \\
& + &   A_2 (z,t)
\exp(-i\omega_2 t+i k_2 z) \\
 &+ & A_3(z,t) \exp(-i\omega_3 t+ik_3z) +c.c. \left. \right\} \; ,
\end{eqnarray}
 the evolution equations of the slowly-varying envelopes
$A_l(z,t)$ ($l=1,2,3$) can be derived, in the limit of weak nonlinearity and
quasi-monochromatic wave trains, by a multiple-scales asymptotic expansion analysis (see, for
instance, \cite{Bang97}). The resulting equations read \cite{Longhi02}:
\begin{subequations}
\begin{eqnarray}
2ik_1 \frac{\partial A_1}{\partial z}=\left[ k_{1}^{2}-k^2(\omega_1+i \partial_t)
\right] A_1-\frac{2k_{1}^{2}}{n_{1}^{2}}d_{eff}A_{2}^{*}A_3 \;\;\;\;\;\; \\
2ik_2 \frac{\partial A_2}{\partial z}=\left[ k_{2}^{2}-k^2(\omega_2+i \partial_t)
\right] A_2-\frac{2k_{2}^{2}}{n_{2}^{2}}d_{eff}A_{1}^{*}A_3 \;\;\;\;\;\; \\
2ik_3 \frac{\partial A_3}{\partial z}=\left[ k_{3}^{2}-k^2(\omega_3+i \partial_t)
\right] A_3-\frac{2k_{3}^{2}}{n_{3}^{2}}d_{eff}^{*} A_1A_2 \;\;\;\;\;\;
\end{eqnarray}
\end{subequations}
where $n_l=n(\omega_l)$ ($l=1,2,3$) are the refractive indices at the three carrier wavelengths,
\begin{equation}
d_{eff}(z) \equiv \frac{1}{2} \overline {\chi^{(2)}(z) \exp(i \Delta k z) }=
\frac{1}{2} \chi^{(2)}_{1}(z),
\end{equation}
is the effective nonlinear interaction coefficient for first-order QPM, and the overline denotes a spatial average over a few modulation periods of the QPM
grating. For a square-wave (+/-) modulation of the ferroelectric domains with 50$\%$ duty cycle and uniform period, one has 
\begin{equation}
d_{eff}(z)=\frac{2} {\pi} d_0 W(z),
\end{equation}
 where $d_0$ is the nonlinear coefficient in the absence of the grating and the real envelope $W$, with $0 \leq W(z) \leq 1$, can be tailored rather arbitrarily with the methods demonstrated in Ref.\cite{QPM}, for example by means of the domain cancellation technique.\\ The linear operators on the right hand side of Eqs.(5) describe the
linear dispersive and absorptive properties of the medium at {\it any} order of approximation. In the following, we will consider spectral
regions of transparency for the medium, so that we will neglect the imaginary part of
 $k(\omega)$. In addition, we will assume a strong and continuous-wave pump field, so that $A_2$ can be taken to be constant (independent of space and time) in Eqs.(5a) and (5c). Under the no-pump-depletion approximation, one can thus write
 \begin{subequations}
 \begin{eqnarray}
2ik_1 \frac{\partial A_1}{\partial z}=\left[ k_{1}^{2}-k^2(\omega_1+i \partial_t)
\right] A_1-\frac{2k_{1}^{2}}{n_{1}^{2}}d_{eff}A_{2}^{*}A_3 \;\;\;\;\;\; \\
2ik_3 \frac{\partial A_3}{\partial z}=\left[ k_{3}^{2}-k^2(\omega_3+i \partial_t)
\right] A_3-\frac{2k_{3}^{2}}{n_{3}^{2}}d_{eff}^{*} A_1A_2 \;\;\;\;\;\;
\end{eqnarray}
\end{subequations}
 
\subsection{Driven two-level atom analogy}
The coupled equations (8a) and (8b) describing the SFG process in the no-pump-depletion approximation, when written for monochromatic waves, bear a close analogy to the optical Bloch equations describing the dynamics of a two-level atomic system driven by a near-resonant optical pulse. Such an analogy, which is fruitful for the prediction of the transparency effect presented in the next section, has been previously discussed in the monochromatic case in Ref.\cite{an1} and applied to efficient broadband SFG based on the analogue of rapid adiabatic passage using chirped QPM gratings \cite{an1,an2}. Other analogies between multi-step frequency conversion processes in nonlinear second-order optical media and coherent population transfer in coherently-driven multi-level atomic systems, including stimulated Raman adiabatic passage, have been highlighted in the recent literature as well \cite{an3,an4}.\\
To show the equivalence of Eqs.(8a) and (8b) with the optical Bloch equations of a driven two-level  atom \cite{an1,an4}, let us consider a monochromatic signal  wave with frequency offset $\Omega$ from the reference frequency $\omega_1$. Since Eqs.(8a) and (8b) are linear ones, the general case of an incident pulsed wave is obtained by standard Fourier analysis starting from the solution of the monochromatic case. After setting in Eqs.(8a) and (8b)
\begin{subequations}
\begin{eqnarray}
A_1(z,t) & = & u(z) \exp[-i \Omega t+i \beta(\Omega) z] \\
A_2(z,t) & = & \frac{n_1}{n_3} \sqrt{\frac{k_3}{k_1}} v(z) \exp[-i \Omega t+i \beta(\Omega) z] 
\end{eqnarray} 
\end{subequations}
with
\begin{equation}
\beta(\Omega) \equiv -\frac{k_1^2-k^2(\omega_1+\Omega)}{4 k_1}-\frac{k_3^2-k^2(\omega_3+\Omega)}{4 k_3}
\end{equation}
one obtains the following coupled equations for the amplitudes $u(z)$ and $v(z)$
\begin{subequations}
\begin{eqnarray}
i \frac{du}{dz} & = - \delta u -q(z) v \\
i \frac{dv}{dz} & =  \delta v -q^*(z) u 
\end{eqnarray}
\end{subequations}
where we have set
\begin{eqnarray}
q(z) & = & \frac{\sqrt{k_1 k_3}A_2^* d_{eff}(z)}{n_1n_3} \\
\delta & = & \delta(\Omega)=\frac{k_3^2-k^2(\omega_3+\Omega)}{4 k_3}-\frac{k_1^2-k^2(\omega_1+\Omega)}{4 k_1}. \;\;\;\;\;
\end{eqnarray}
Equations (11) are analogous to the optical Bloch equations for a driven two-level atom describing the transition between the two atomic levels induced by a nearly resonant optical pulse with Rabi frequency $q(z)$ and frequency detuning $2 \delta$ [Fig.1(b)] \cite{an1}. Note that the detuning $\delta$, as given by Eq.(13), accounts for material dispersion at any order. A simple expression of $\delta(\Omega)$ is obtained when group velocity dispersion (and higher-order dispersion effects) are negligible, and $k^2(\omega_l + \Omega)$ can be expanded in power series up to first order in $\Omega$. After setting $k^2(\omega_l+\Omega) \simeq k_l^2+2 k_l (\omega_l) \Omega / v_{gl}$, where $v_{gl}=(d k / d \omega)_{\omega_l}^{-1}$ is the group velocity at the carrier frequency $\omega_{l}$, one simply obtains
\begin{equation}
\delta(\Omega) \simeq \frac{\Omega}{2} \left( \frac{1}{v_{g1}}-\frac{1}{v_{g3}} \right)
\end{equation}
where $v_{g1}$ and $v_{g3}$ are the group velocities of signal and SFG waves, respectively. In the following analysis, we will assume that the QPM grating is not chirped, so that the Rabi frequency $q(z)$ entering in Eqs.(11) can be assumed to be real.\\
As shown in the next section, the broadband transparency effect predicted in this work is based on the two-level atom analogy and existence of off-resonance Rabi pulses $q(z)$, which do not transfer population between the two atomic levels.  It should be noted that the two-level atom analogy can be established for the SFG process, but not for other second-order nonlinear interactions like parametric amplification involving a down-conversion process. In the latter case, which was considered in \cite{referee1, Longhi02}, the underlying equations for idler and signal waves differ from Eqs.(11) because of the replacement $ - q^*(z)u \rightarrow q^*(z)u$ on the right hand side of Eq.(11b). The resulting coupled equations describe an  exponential (rather than oscillatory) behavior of interacting waves, and are similar to coupled-mode equations found in Bragg scattering theory of counter-propagating waves \cite{referee1,Longhi02}. As a result, broadband transparency effects are prevented in parametric amplification, where only narrow transparency windows can be opened in the spectral gain curve and associated to superluminal group velocities \cite{referee1,Longhi02}. 

\section{Transparency in sum frequency generation}
\subsection{Theoretical analysis}
The solution to Eqs.(11), from the input $z=-L/2$ to the output $z=L/2$ planes of the nonlinear medium, can be written in the general form
\begin{equation}
\left(
\begin{array}{c}
u(L/2) \\
v(L/2)
\end{array}
\right)=
\left(
\begin{array}{cc}
\mathcal{M}_{11}(\delta) & \mathcal{M}_{12}(\delta) \\
\mathcal{M}_{21}(\delta) & \mathcal{M}_{22}(\delta)
\end{array}
\right) \times \left(
\begin{array}{c}
u(-L/2) \\
v(-L/2)
\end{array}
\right)
\end{equation}
where the transfer matrix $\mathcal{M}$ is unimodular with $\mathcal{M}_{11}=\mathcal{M}_{22}^*$, $\mathcal{M}_{21}=-\mathcal{M}_{12}^*$ and $|\mathcal{M}_{11}|^2+|\mathcal{M}_{12}|^2=1$. For signal excitation at the input plane $z=-L/2$, i.e. for $v(-L/2)=0$, the spectral transmission of the signal wave is simply given by
\begin{equation}
t(\Omega)=\left( \frac{u(L/2)}{u(-L/2)}\right)_{v(-L/2)=0} =\mathcal{M}_{11}.
\end{equation} 
Note that $t(\Omega)$ can be factorized as
\begin{equation}
t(\Omega)=t_0(\Omega) \exp[i \delta(\Omega)L],
\end{equation}
where $\exp[i \delta(\Omega)L]$ is the spectral transmission (phase delay) introduced by the medium in the absence of the nonlinearity, i.e. for $q(z) \equiv 0$, and $t_0(\Omega)$ accounts for the   nonlinear interaction. {\it Broadband transparency} is realized provided that 
\begin{equation}
\mathcal{M}_{12}(\delta)=\mathcal{M}_{21}(\delta)=0 \; ,\;\; |\mathcal{M}_{11}(\delta)|=|\mathcal{M}_{22}(\delta)|=1
\end{equation}
for any detuning $\delta$, i.e. $|t_0(\Omega)|=1$ for any frequency offset $\Omega$ of the carrier wave from the reference frequency $\omega_1$. This means that for an {\it arbitrary} optical signal pulse propagating into the nonlinear medium no SFG field is produced at the output of the crystal in spite of phase matching. We note that {\it invisibility} is a more stringent condition than transparency, since it requires $t_0(\Omega)=1$ for any frequency $\Omega$. If the nonlinear medium is transparent but the phase of $t_0(\Omega)$ is not flat, a propagating signal pulse in the medium would suffer for an additional phase delay arising from the nonlinear interaction, resulting in pulse distortion as compared to the invisible regime $\chi^{(2)}=0$ of linear propagation.\\  
A {\it necessary} condition for the observation of transparency can be readily established as follows. Exact solution to the Bloch equations (11) is available at exact resonance $\delta=0$ for an arbitrary shape of the Rabi frequency $q(z)$. In fact, for $\delta=0$ one simply has $\mathcal{M}_{11}= \cos \mathcal{A} $, where
\begin{equation}
\mathcal{A}=\int_{-L/2}^{L/2} q(z) dz
\end{equation}
is the 'area' of the driving pulse in the quantum mechanical analogy. Hence transparency at $\delta=0$ requires 
 \begin{equation}
 \mathcal{A}= N \pi
\end{equation}
with $N$ integer. Equation (20) provides a {\it necessary} condition  for broadband transparency, since it ensures transparency at resonance $\delta=\Omega=0$. However, for a general profile $q(z)$ transparency is not found far from resonance, i.e. for $\delta \neq 0$, especially if there is a non-neglibile group velocity mismatch between  signal and sum-frequency waves [see Eq.(14)]. For example, let us consider the simplest case $W(z)=1$, corresponding to a non-apodized (uniform) QPM grating, so that $q(z)=q_0$ constant in the range $-L/2<z<L/2$. According to Eq.(20), transparency at resonance $\delta=0$ requires $q_0=N \pi/L$. However, for $\delta \neq 0$ the transmittance in not unity. In fact, after a simple calculation one finds
\begin{equation}
|t_0(\delta)|^2=\cos^2 \left( \sqrt{q_0^2+\delta^2} L \right) +\frac{\delta^2}{\delta^2+q_0^2} \sin^2 \left( \sqrt{q_0^2+\delta^2} L  \right).
\end{equation}
 The question thus arises whether there exist special profiles $q(z)$ such that $|t_0(\delta)|^2=1$ for any vaue of the detuning $\delta$. In the theory of driven two-level atoms, it is known \cite{Aku} that for $q(z)$ of the form 
\begin{equation}
q(z)=\frac{q_0}{\cosh( \alpha z)},
\end{equation}
transparency at any value of detuning $\delta$ can be realized whenever the condition (20) on the area is satisfied. The parameter $\alpha$ entering in Eq.(22) can be taken arbitrarily, and its inverse $ 1/ \alpha$ basically determines the characteristic length of nonlinear interaction.
For such a profile of $q(z)$, exact solution for the optical Bloch equations (11) can be obtained in terms of hypergeometric functions \cite{Aku}, and the transparency at special values of the amplitudes $q_0$ satisfying the area condition (20) can be explained in term of supersymmetric quantum mechanics \cite{Aku2}. Assuming a medium length $L$ such that $\cosh(\alpha L/2) \gg 1$, the nonlinear correction $t_0(\Omega)$ to the transmission coefficient $t$ can be obtained in a closed form and reads explicitly
\begin{equation}
t_0(\Omega)=\frac{\Gamma(1/2+i \Delta) \Gamma(1/2+i \Delta) }{\Gamma(1/2+i \Delta- \mathcal{A} / \pi) \Gamma(1/2+i \Delta+ \mathcal{A} / \pi)}
\end{equation}
where we have set
\begin{equation}
\Delta \equiv \frac{\delta( \Omega)}{\alpha} \simeq \frac{\Omega}{2 \alpha} \left( \frac{1}{v_{g1}}-\frac{1}{v_{g3}} \right), 
\end{equation}
$\mathcal{A}= \pi q_0/ \alpha$ is the area (defined by Eq.(19) with $L \rightarrow \infty$), and $\Gamma(.)$ is the Gamma function. From Eq.(23) the spectral transmittance $T(\Omega)=|t(\Omega)|^2=|t_0(\Omega)|^2$ for the signal wave can be calculated, which reads
\begin{equation}
T(\Omega)=1- \frac{\sin^2 \mathcal{A}}{\cosh^2 ( \pi \Delta)}
\end{equation}
where $\Delta=\Delta(\Omega)$ is given by Eq.(24). 
Note that broadband transparency $T=1$ is obtained provided that $\mathcal{A}= N \pi$ with $N$ integer, according to Eq.(20). Once the normalized spatial profile $W(z)$ of the QPM grating is designed according to $W(z)=1/ \cosh( \alpha z)$, from Eqs.(7) and (12) it follows that the transparency condition $\mathcal{A}= N \pi$ is met for special values of the pump amplitude $A_2$. In terms of the intensity $I_2=(1/2) \epsilon_0 c_0 n_2 |A_2|^2$ of the strong pump wave, the transparency condition is satisfied provided that $I_2= N I_{tr}$, where $N$ is an integer number and $I_{tr}$ is the transparency pump intensity given by
\begin{equation}
I_{tr}=\frac{1}{32} \frac{\epsilon_0 c_0 n_1 n_2 n_3 \lambda_1 \lambda_3 \alpha^2}{d_0^2}
\end{equation} 
For a pump intensity $I_2=NI_{tr}$, the nonlinear medium is broadband transparent, i.e. no SFG wave is generated at the output of the crystal for any arbitrarily broadband incident signal pulse. In fact, once the area condition (20) is satisfied the transparency bandwidth is in principle infinite according to Eq.(25). In practice, however, deviations of the profile of the effective susceptibility from the ideal sech shape or pump intensity deviations from the transparency value  result in the appearance of a spectral region around the phase matching condition where the transmittance is not unitary. For example, if the pump intensity $I_{2}$ is close to but slightly detuned from the transparency value $I_{tr}$,  according to Eq.(25) 
 transparency is not observed in a spectral region with a bandwidth $\Delta \Omega$ determined by the condition $\pi | \Delta| \sim 1$, i.e. by the group velocity mismatch and interaction length $\Delta \Omega \sim (2 \alpha / \pi) |1/v_{g1}-1/v_{g3}|^{-1}$ [see Eq.(24)].\\ 
 It should be noted that, even thought the transparency condition is met, the nonlinearity of the medium {\it is not invisible} since the phase of $t_0(\Omega)$, as given by Eq.(23), is not flat. For example, in the simplest case $N=1$, i.e. for 
$\mathcal{A}= \pi$, one has
\begin{equation}
t_0(\Omega)=\frac{\delta+i \alpha/2}{\delta-i \alpha/2}=\exp[i \phi(\Omega)]
\end{equation}
with 
\begin{equation}
\phi(\Omega)=2 {\rm atan} \left( \frac{\alpha}{ 2 \delta( \Omega) }\right) \simeq 2 {\rm atan} \left[ \frac{\alpha v_{g1} v_{g3} }{ \Omega(v_{g3}-v_{g1})} \right].
\end{equation}
The additional phase delay $\phi(\Omega)$ leads to an effective non-linear induced contribution to the linear material dispersion, and can be detected in pulse transmission experiments, as discussed in the next subsection.

\begin{figure}
\includegraphics[scale=0.35]{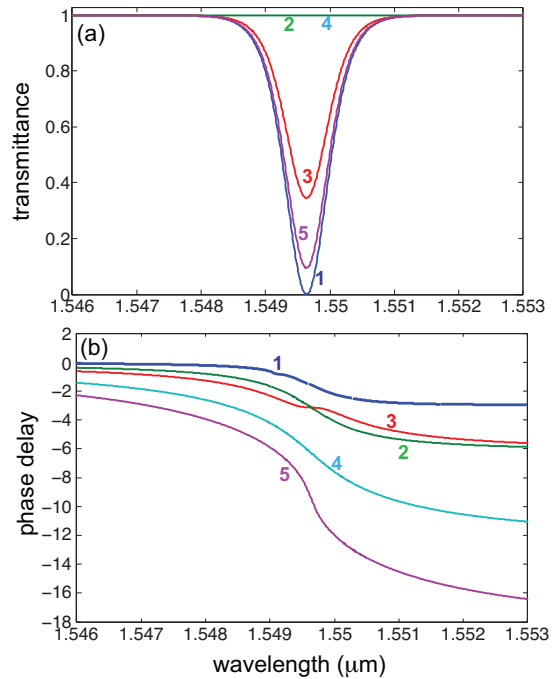}
\caption{(Color online) Numerically-computed (a) spectral transmittance, and (b) phase delay for the signal wave in a 2.5-cm-long PPLN crystal with $\cosh^{-1}(\alpha z)$-apodized profile for $\alpha=5$ cm$^{-1}$ and for increasing values of the pump intensity $I_2$. Curve 1: $I_2=0.5 I_{tr}$; curve 2: $I_2= I_{tr}$; curve 3: $I_2=1.3 I_{tr}$; curve 4: $I_2=2 I_{tr}$; curve 5: $I_2=2.6 I_{tr}$. The pump intensity at transparency is $I_{tr} \simeq 24.37$ MW/cm$^2$.}
\end{figure}

\begin{figure}
\includegraphics[scale=0.35]{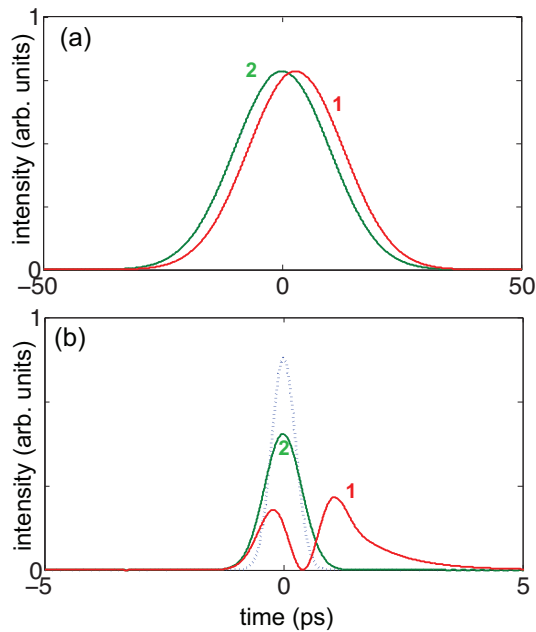}
\caption{(Color online) Numerically-computed propagation of a Gaussian signal pulse at carrier wavelength $\lambda_1=1.55 \; \mu$m in a 2.5-cm-long PPLN crystal with $\cosh^{-1}(\alpha z)$-apodized profile for $\alpha=5$ cm$^{-1}$ and for a FWHM pulse width (a) $\Delta \tau_p=23.5$ ps, and (b) $\Delta \tau_p=589$ fs. Curve 1 shows the transmitted pulse intensity distribution of the signal field for a continuous-wave pump intensity $I_2=I_{tr} \simeq 24.37$ MW/cm$^2$, whereas curve 2 is the transmitted pulse distribution of the signal waveform when the pump field is switched off ($I_2=0$, linear propagation regime). The thin dotted curve [almost overlapped with curve 2 in (a)] is the pulse intensity distribution of the weak Gaussian signal pulse at the input plane of the crystal.}
\end{figure}

\subsection{Numerical results}

To illustrate the phenomenon of SFG transparency and to provide some design parameters, let us consider as an example nonlinear frequency conversion in a periodically-poled lithium niobate (PPLN) crystal pumped at the wavelength $\lambda_2=810$ nm and probed with a weak signal field
at $\lambda_1=1.55 \; \mu$m. The SFG wave corresponds to the wavelength $\lambda_3=532$ nm. We assume extraordinary wave propagation, corresponding to a nonlinear coefficient $d_0=d_{33} \simeq 27$ pm/V. The temperature-dependent dispersion relation $k=k(\omega)=n(\omega)
\omega / c_0$ for extraordinary waves in lithium niobate is determined using Sellmeier
equations from Ref.\cite{Edwards84}. At 25$ ^o$C, one can estimate $n_1=2.1381$, $n_2=2.1748$, $n_3=2.2343$, the group velocities
$v_{g1} \simeq 0.4581 c_0$, $v_{g3} \simeq 0.4069 c_0$, and a first-order QPM grating with period $\Lambda=2 \pi / \Delta k=7.38 \; \mu$m, which is accessible with current poling technology.
As an example, Fig.2 shows the numerically-computed transmittance (modulus square of $t_0$) and phase delay (phase of $t_0$) versus wavelength in a $L=2.4$ cm long PPLN crystal for $\alpha=5$ cm$^{-1}$ and for increasing values of the pump intensity $I_2$. The intensity at transparency is given by $I_{tr} \simeq 24.37$ MW/cm$^2$ according to Eq.(26). Note that, for a non-integer value of the normalized pump intensity $I_2/I_{tr}$, SFG is observed in a wavelength range of the signal wave corresponding to phase matching of the nonlinear interaction. This is shown by curves 1,3, and 5 in Fig.2(a), where the spectral transmittance shows a dip near the wavelength of perfect phase matching. The central wavelength of the dip and its width are determined by the phase matching condition in the nonlinear interaction, i.e. by the QPM grating period, the decay length $1/ \alpha$ of the QPM grating, and the material dispersion. 
As the ratio $I_2 / I_{tr}$ is an integer number (curves 2 and 4 in Fig.2), there is no SFG wave, i.e. the medium in transparent for the signal wave according to the theoretical analysis. Nevertheless, a wavelength-dependent phase delay is accumulated in the nonlinear interaction [see Fig.2(b)], corresponding to an additional non-linear-induced dispersion term for the signal field. In an experiment, the effect of the nonlinear-induced dispersion at the transparency regime can be detected by  comparing the propagation of a short optical pulse along the medium with the pump field switched off and on.  This is illustrated in Fig.3, which shows the numerically-computed propagation of a Gaussian input signal pulse $A_2(-L/2,t) \propto \exp[-(t/ \tau_p)^2]$ along the 2.4 cm-long PPLN crystal when the pump intensity is tuned at the transparency value $I_2=I_{tr}$ (curve 1) and when it is switched off $I_2=0$ (curve 2). The pulse duration $\Delta \tau_p$, defined as the full-width at half maximum of the field intensity, is related to $\tau_p$ by the relation $\Delta \tau_p=\sqrt{2 {\rm ln} 2} \tau_p$. For a relatively long input pulse [Fig.3(a), $\Delta \tau_p \simeq 23.5$ ps], the linear dispersion of the medium is negligible, and the nonlinear-induced dispersion is responsible for a time delay of the transmitted pulse, given by the group delay $\tau_d=d (\phi / d \Omega)=2(v_{g1}-v_{g3})/(\alpha v_{g1}v_{g3}) \simeq 3.6$ ps. For shorter pulses [Fig.3(b), $\Delta \tau_p \simeq 589$ fs], the linear dispersion of the medium is non-negligible [curve 2 in Fig.3(b)], and the additional dispersion introduced by the non-linear interaction at $I_2=I_{tr}$ is responsible for strong pulse reshaping. In particular, one can observe pulse splitting with a long pulse tail [curve 1 in Fig.3(b)].

\section{Conclusions and outlook}
Optical waves propagating in a  linear but inhomogeneous medium generally show reflection and scattering when the material properties rapidly change over a distance of the order of the optical wavelength \cite{r7bis}. However, proper tailoring of the dielectric permittivity can suppress scattering and the medium thus appears to be transparent \cite{r5,Kay,coat}. A different kind of 'scattering' can occur in the frequency domain when the optical waves propagate in a nonlinear $\chi^{(2)}$ medium. When phase matching conditions are met, efficient frequency conversion can occur, and an optical wave at a reference frequency (signal field) can be amplified or attenuated owing to frequency conversion.  Here we have investigated the possibility to realize optical transparency in the process of sum frequency generation in a second-order nonlinear crystal. By exploiting the quantum-optical analogy between the process of SFG in the undepleted pump approximation and the coherent excitation of a two-level atom by a near-resonant pulse with tailored shape and pulse area, we have shown that broadband transparency can be realized in the nonlinear crystal with an engineered QPM grating. Such a result could be of interest in the nonlinear control of material transparency and is expected to motivate further theoretical and experimental studies in the field of transparency and invisibility in nonlinear media. A few natural extensions to the present study can be envisaged. For example, is it possible to engineer the nonlinear interaction to realize {\it one-way} transparency? Also, can one tailor the nonlinearity of the medium to make it invisible? One possibility might be to engineer the material properties to allow for an {\it imaginary} part of the nonlinear susceptibility \cite{uff1}, i.e. to explore the full domain of {\it complex} non-linear susceptibility. In this case, transferring the recent proposal by Horsley and coworkers  \cite{r5}  of spatial Kramers-Kronig relations for linear susceptibilities to the nonlinear ones, it would be possible to realize one-way transparency and non-linear invisibility. Other extensions of the present study might be the analysis of transparency and invisibility in two-dimensional QPM gratings, in nonlinear interactions with phase-matched counter-propagating waves \cite{uff2}, and in frequency wave mixing based on third-order nonlinear media.

RESPONSE TO YOUR QUERIES

Q1. There are no funding institutions/organization to acknowledge for this article.

Lines 8-9: In the sentence: ?? and oscillation provide a powerful tools for coherent ??, please write ?provide powerful tools? in place of  ?provide a powerful tools? (i.e. delete ?a?  after ?provide?).

Lines 26, 27: In the sentence: ?? nonlinear crystals, such frequency doubling, ?.? please write ?such as? in place of ?such? (i.e. please add ?as?after ?such?). 

In Eq.(5), first term $2g(z)$ on the left hand side, please write $g$ using italic style 

Line F5:5, in the second  equation of the line $2 gl=26?.$: the space between ?2? and ?g? should be deleted, and ?g? should be written using italic style. The same holds in line 265 (second equation in the line).
 
Line 282, the sentence: ? The ?upconversion? OPO provides ?? please write ?The ?upconversion? parametric amplification provides??, i.e. please write ?parametric amplification? in place of ?OPO?.

\end{document}